\documentstyle[11pt,fleqn,epsfig]{article}     
\newcommand{\hw}{D$_2$O}
\newcommand{\lw}{H$_2$O}
\newcommand{\beq}{\begin{equation}}
\newcommand{\eeq}{\end{equation}}
\begin{document}

\noindent

\begin{center}
\large
An Early History of Heavy Water

\normalsize

Chris Waltham

Department of Physics and Astronomy, University of British
Columbia

Vancouver B.C., Canada V6T 1Z1

{\it cew@phas.ubc.ca}

\vspace*{3mm}

August 1998; revised June 2002; minor corrections October 2011

\end{center}

\section*{Abstract}

Since 1945 Canada has had a nuclear power industry based on
reactor designs that use natural uranium and heavy water. 
The tortuous and improbable sequence of events which led to
this situation
is examined.

\newpage

\section{Introduction}

Pure heavy water, \hw , is the oxide of the heavy stable isotope of
hydrogen, deuterium, denoted by the symbols $^2$H or D. It is physically
and chemically almost identical to ordinary ``light" water, \lw , except
in that its density is 10\% higher, hence its name. The answer to the
question everyone asks is ``Yes, you can drink it"; the Material Safety
and Data Sheet states that dizziness and nausea can occur if you ingest
10\% of your body mass of heavy water, and LD$_{50}$ is 30\% of body mass.
Hence it is safer than ethanol.

Most of the deuterium (heavy hydrogen) was formed about 10 minutes after
the Big Bang, along with other very light isotopes presently found in the
universe. More recently, 2.5 billion years ago, most of the deuterium
atoms on the earth were incorporated into water molecules.  As a small
isotopic fraction of natural hydrogen (0.015\%), deuterium existed then,
as
now, mostly in the form of HDO molecules. And this is the way things
remained until deuterium, and thus heavy water was discovered in 1931.
However, before we examine how deuterium was discovered, let us examine
the events which led to the manufacture of heavy water, for these seeds
were sown before those of discovery.

\section{Birkeland and Norsk Hydro}

At the turn of this century Professor Kristian Birkeland\cite{1} of the
University of Kristiania (soon to be renamed Oslo) was experimenting with
a device called the Terella. This was a laboratory model of the earth,
complete with magnetic field, placed in an evacuated vessel into which
ions could be injected at high voltage. By observing the discharge glow,
Birkeland was able to understand the three-dimensional structure of the
Aurora Borealis. However, the equipment was expensive and money, then as
now, was hard to come by. Kelvin suggested to Birkeland that research into
armaments might prove lucrative and allow him to continue with the
Terella; this was not the first time such an idea had occurred to a
physicist. Thus Birkeland with his expertise in electromagnetism, set about
producing an {\it elektriske kanon}, a rail-gun.

On Feb. 6th, 1902 Birkeland's {\it kanon} short-circuited and exploded
during a
test, but the disappointment he felt was muted by the observation of a
disc-shaped arc, spread by the magnetic field, and the smell of nitrogen
oxides. The reason this was intriguing was that the world was then gripped
by a fear greater even than that of the looming conflict in Central Europe
- that of world famine. Chilean nitrate deposits, on which the world
depended as a fertilizer, were on the brink of running out, and chemists
were scurrying to find an economical way of fixing atmospheric nitrogen.
Clearly the acrid brown stench of nitrogen oxides was the smell of
nitrogen being fixed, as it was a short hop skip and a jump to nitric acid
and any nitrate you please.

Birkeland was not the first to fix nitrogen by electric arc - Crookes in
Manchester already had a pilot plant producing calcium nitrate by this
means - but the disc-shaped arc promised a high yield\cite{2}. A
week later Birkeland met a man who was already working on the means
which would make economical production possible. That man was Sam Eyde, a
civil engineer who was fascinated by the enormous potential of Norway's
mountains and rainfall for the production of hydro-electricity. After that
things moved with breath-taking speed. A week after meeting, Eyde and
Birkeland submitted a patent for artificial fertilizer. They obtained
money from Swedish financiers, the Wallenbergs, and a mere three years
later a hydroelectric plant had been built out of the wilderness at
Notodden and a Birkeland-Eyde arc furnace was producing the first
Norgesalpeter - Norwegian Saltpeter, i.e. calcium nitrate.

In the same year as the first Norgesalpeter was produced, Fritz Haber
discovered a much better way of producing nitrates via ammonia in what 
is now known as the Haber process. Indeed, the Norwegians soon abandoned
the arc furnaces and adopted Haber's idea. However, Birkeland's discovery
had started something irreversible: the large scale development of
hydro-electric power in Norway by the company he and Eyde had founded:
Norsk Hydro.

Now we will examine the other chain of events, that of discovery. First,
however I should note the strange and sad circumstances of Birkeland's
death in 1917 - too soon for him to know anything about heavy water. He
had been spending some time in Egypt, where it was hoped that the warm
weather would improve his health. He wished to return home to Norway, but
found the direct route barred by the First World War. During the ensuing
tortuous detour, he died, in Tokyo.

\section{Discovery}

The path that led to the discovery of heavy hydrogen and heavy water was
every bit as tortuous as Birkeland's attempt to get home from Egypt. In
1913 Arthur Lamb and Richard Lee at New York University were trying to
improve measurements of the density of water; this was a very important
quantity to know accurately because of its importance as a
standard\cite{3}. They were attempting an accuracy of 200 ppb, but were
not able to get agreement between samples taken from different
geographical locations to better than 800ppb. They did not grasp the
significance of this discrepancy - the abundance of deuterium varies due
to rates of evaporation and condensation - despite the fact that the
concept of isotopes as developed by Frederick Soddy and Kasimir Fajans was
a significant piece of scientific news that year.

In 1929 William Giauque and Herrick Johnston discovered $^{17}$O and
$^{18}$O. Two years later Harold Urey at Columbia constructed chart of
missing and known isotopes and from gaps in the pattern mused on the
possible existence of $^2$H, $^3$H and $^5$He. In the same year Raymond T.
Birge at Berkeley and Donald H. Menzel at Lick noted\cite{4} that all
spectrograph measurements were based on $^{16}$O=16, but chemical methods
based on natural oxygen=16. However the chemical value for the mass of
hydrogen was 1.00777$\pm$0.00002 on this scale which compared very closely
to Aston's spectroscopic value of 1.00778$\pm$0.00015. Hence, the
existence of heavy isotopes of oxygen must be balanced by one or more
heavy isotopes of hydrogen; if this was all $^2$H, then its abundance
compared to $^1$H must be around 1:4500.

Urey and Murphy set about looking for $^2$H in the spectrum of natural
hydrogen; the expected shift was very small, $m_e/2m_p$, or 1.8
Angstr\"oms 
on
6565 Angstr\"om
Balmer alpha line. Seeing only tantalizing hints of a second line in
natural hydrogen they proceeded to distill liquid hydrogen, hoping to
leave the heavy isotope preferentially behind. Victor LaMer, a Columbia
electrochemist had argued that electrolysis of water would have no effect
on the isotopic fraction; it never occurred to them that the huge
fractional mass difference between H and D would make electrolysis and
excellent separator! In fact it was worse than that because Brickwedde
from NBS who prepared the samples chose nice and pure but
deuterium-depleted electrolytic hydrogen for the distillation process.
Eventually the deuterium concentration in the electrolyte increased with
use and the line appeared. After the errors were discovered an isotopic
ratio of 1 in 6500 was obtained, very close to the 1 in 7000 value
accepted today. Soon afterwards Edward Washburn and Urey\cite{5} used the
electrolytic method to prepare the first isotopically enriched sample of
heavy water.

Urey had trouble getting travel funds to present his work at December 1931
APS meeting in Tulane; an appeal had to made to the President of Columbia
to provide his travel costs. Two years later he was awarded the Nobel
Prize for Chemistry.

As a footnote to Urey's discovery, in 1935 Aston found error in his
measurement of the mass of hydrogen\cite{6}. Now he got 1.00813 on
physical scale or 1.0078 on chemical scale which implied if anything a
lighter isotope of hydrogen. In fact the existence or otherwise of rare
hydrogen isotopes was well buried in the experimental errors. But it was
too late: deuterium had been discovered.

\section{Production}

The first concentration of deuterium in water was achieved by using
electrolysis; the light isotope of hydrogen was preferentially evolved,
leaving behind water enriched in deuterium (see appendix). Separation
factors
of about six can easily be obtained and double this is possible. The
economics are not great, however. It is necessary to electrolyze 2700
litres of natural water to obtain one litre of water enriched in deuterium
by 10\%; this requires 320 MW-hours of electrical energy. By repeating the
process of electrolysis, pure heavy water can be produced; this is a
conceptually simple method but it is enormously expensive.

Scientific interest in heavy water started to emerge soon after its first
production by Lewis and MacDonald\cite{7} as its analytical potential in
chemistry and biology, and its possibilities in cancer therapy, as well as
more direct application in nuclear physics, became apparent.  Small
amounts (grams) for scientific use were available in the United States by
1933.
 
\section{Norsk Hydro Enters the Field}

\begin{figure}
\center{\epsfig{file=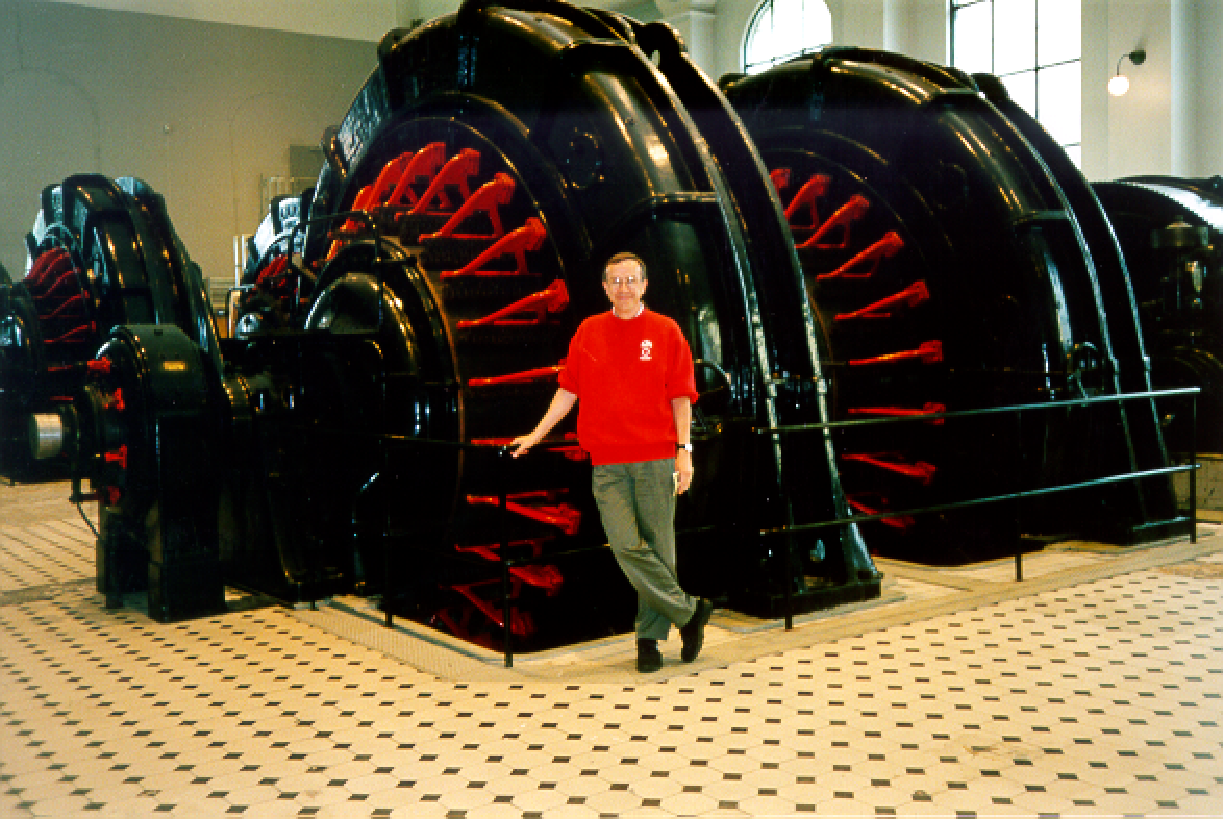,width=15cm}}
\caption{Hydro-electric generator building at Vemork, just outside
Rjukan,
now the Norsk Industriarbeidermuseum (Industrial Workers Museum). The
heavy water plant was in a (rather unattractive) building, now demolished,
in front of this one. (Author photo)}
\label{fig:vemork}
\end{figure}

Jomar Brun, the Head of Hydrogen Research at Norsk Hydro, and Leif
Tronstad, a physicist from Trondheim, realized that the conditions for
large-scale production (kg) of heavy water existed at Norsk Hydro's plant
in Rjukan, where large amounts of water were already being electrolyzed as
part of the Haber-Bosch process for producing ammonia for nitrogen
fertilizer. They drew up a plan, with some involvement from Karl-Friedrich
Bonh\"offer, a German physical chemist at Leipzig (and brother of Dietrich, the theologian), for the industrial
production of heavy water. It was in many ways an astonishing venture as a
large amount of equipment had to be built - hundreds of combined
electrolysis, combustion and condensation cells - and the market must have
been uncertain. However, Norsk Hydro went ahead, and built a plant by the
generator building at Vemork, just outside Rjukan (fig.\ref{fig:vemork}).

A glance at the first order book (fig.\ref{fig:order}) shows impressive
and rapid progress. The first order of one litre of 0.43\% heavy water was
shipped to Birkbeck College, London, on August 9th, 1934, and by
mid-November 95\% was achieved (and sent, not unreasonably, to Tronstad's
lab in Trondheim). By January 1935, the first production of more than 99\%
pure heavy water was available at 10Kr (about 50 cents) per gram, a tenth
of the American price. Tronstad and Brun themselves did important
standards work on deuterium and heavy water\cite{8}.

About this time, the idea of producing heavy water at Trail, British
Columbia, was first discussed in correspondence between the Canadian
National Research Council and the owners of the Trail hydroelectric plant,
Cominco\cite{9}. This plant was the largest producer of electrolytic
hydrogen in North America; it was used, as was Norsk Hydro's, for making
ammonia for the fertilizer and explosives industry.

\section{Loss of Innocence}

For a few years life with deuterium and heavy water was pleasant and
productive. As a target for nuclear physicists to play with deuterium
provided an excellent source of neutrons, then a primary tool in
fundamental subatomic physics. Paul Harteck (whom we shall meet again), a
young German working with Rutherford and Marc Oliphant at the Cavendish
Laboratory used it as a stepping stone to discover an even heavier
hydrogen isotope, tritium\cite{10}. And although heavy water never panned
out as a cure for cancer (it was not possible to utilize the different
cell growth rates in heavy and ``light" water to this effect), biologists
made hay with the material; for example George Hevesy in a classic set of
experiments used it to establish the fraction of water in mammalian
biology\cite{11}.
  
In 1937, Hans von Halban and Otto Frisch, working at Bohr's Copenhagen
laboratory, noticed that heavy water had very low neutron absorption
compared to light water. Protons and neutrons like to be paired off; the
cross section for neutron capture on a proton to produce a deuteron is
much greater than that for neutron capture on the already paired deuteron.
This latter event does occasionally happen, producing $^3$He, and this in
turn loves to capture neutrons to produce the doubly-paired $^4$He.

Around Christmas of 1938 the world changed when Otto Hahn and Fritz
Strassman at the Kaiser Wilhelm Institute in Berlin discovered that low
energy (thermal) neutrons could split uranium nuclei, releasing an
enormous amount of energy. Using the liquid-drop model, Hahn's erstwhile
colleague Lise Meitner and her nephew Otto Frisch were able to understand
the basics of this fission process and deduced the energy release to be
about 200 MeV, or about 10 million times larger than the energy change in
a typical chemical reaction\cite{12}. In January 1939 Hahn and
Strassman\cite{13}
suggested that thermal neutron-induced fission of uranium could release
secondary neutrons. These could go on to produce a chain reaction provided
they were slowed down (moderated) to thermal energies to increase their
chance of causing further fission (see Appendix).

By April, Fred\'er\'eric Joliot and his colleagues Hans von Halban and Lew
Kowarski from the Coll\`ege de France in Paris had observed these
secondary neutrons, and they measured the number produced in each
fission\cite{14}. By August they found that blocks of uranium oxide showed
increased fission activity when immersed in ordinary water\cite{15}.
However, absorption of neutrons on hydrogen prevented a self-sustaining
chain reaction.

Across the Atlantic, Fermi and Szilard at Columbia University examined
alternative moderators and quickly decided on ultra-pure graphite. Work on
carbon was also begun by George Laurence at the Canadian National Research
Council in Ottawa (later joined by Bern W. Sargent from Queen's
University)\cite{16}. Szilard started to persuade all physicists working
on fission to cease publishing; he sent a cable to Joliot on April 6th
requesting a delay in further publications ``in view of possible misuse in
Europe"\cite{17}.

Sometime in early summer, the Paris team alighted on the idea of using
heavy water\cite{15} as a moderator. Deuterium was known to have a much
lower absorption cross section for neutrons than ordinary
hydrogen\cite{18} and its low mass makes it an almost ideal moderator. Use
of heavy water would thus make a self-sustaining chain reaction more
accessible. Halban and Kowarski did some simple modelling of neutron
moderation and this was enough to suggest \hw\ as the best
candidate\cite{15}. The putative moderating qualities of deuterium were
then widely appreciated; as early as the beginning of February 1939
Oppenheimer had a crude drawing on his blackboard of an atomic bomb made
of uranium deuteride\cite{19}.

At the end of October Halban, Joliot and Kowarski deposited a sealed
envelope with the Academy of Science. This was opened in 1949 and the
paper inside ``Sur la possibilit\'e de produire dans un milieu uranif\`ere
des r\'eactions nucl\'eaires en chaine illimit\'ee'' was published in
Comptes Rendus\cite{20}. The paper shows the group had a very firm
theoretical grasp of reactor physics and includes what we now know as the
Fermi four-factor formula. When much later Patrick Blackett wrote Joliot's
obituary he said ``There is little doubt that, had the war not intervened,
the world's first self-sustaining chain reaction would have been achieved
in France"\cite{21}.

We now know a natural uranium reactor is possible with only three
practical moderators: heavy water, ultra-pure graphite, and beryllium.
Heavy water is by far the best, but in 1939 it was not available in very
large quantities. Graphite was more convenient, being very common, but it
had to be extremely pure. The purity of a graphite moderator is crucial as
natural impurities tend to be highly efficient neutron absorbers. Early
measurements of absorption cross-sections made by Joliot and independently
by Walter Bothe and Peter Jensen\cite{22} in Heidelberg were much too
large for this reason. Thus carbon was rejected in Europe and this
precipitated a fight for the Norwegian heavy water.

After moving to Chicago, Enrico Fermi used 40 tons of uranium and uranium
oxide and 385 tons of ultra-pure graphite to achieve the first
self-sustaining chain reaction, on 2nd December 1942.

\section{La Bataille de L'Eau Lourde\cite{23}}

As the commercial and military potential of heavy water sank in, French
military intelligence (the Deuxi\`eme Bureau) learned that there was
considerable German interest in not only obtaining existing Norwegian
stocks, but in a contract for large and regular supplies\cite{24,25}. In
March 1940, Lieutenant Allier of the Deuxi\`eme Bureau left Paris for Oslo
to negotiate with Norsk Hydro.  The resulting agreement ensured that
France was to have not only the 185kg of heavy water then at Rjukan,
immediately, but also a priority claim to the plant's entire output.
Allier suspected he was a target for German agents, and took the
precaution of double-booking himself and his cargo on both a flight to
Scotland, and on one to Amsterdam.  It seems his fears were justified, as
Luftwaffe aircraft forced the Amsterdam flight to land in Hamburg, where
it was thoroughly examined. Allier and his 26 cans of heavy water landed
safely in Scotland; then he travelled to the French Military Mission in
London, and eventually across the Channel. The heavy water was installed
in a special air raid shelter in the Coll\`ege de France.

In the summer of 1940, as France faced defeat, Dautry, the French
Armaments Minister, ordered Joliot to ensure that his cans of heavy water
did not fall into enemy hands.  Hans von Halban, his colleague, first took
the cargo to Mont-Dore, a spa town in central France. He put his wife and
one-year-old daughter in the front of the car, one gram of Marie Curie's
radium in the back, and, to minimize any possible danger from radiation,
the cans of heavy water in between. Upon arrival, Halban was allowed to
lodge the cans in the safety of the town's women's prison. The following
morning, after they had been moved for safety to the condemned cell of the
prison in nearby Riom, Halban began to set up a new laboratory in a small
villa. Shortly afterwards, the evacuation order came. A number of
prisoners serving life sentences were ordered to move the world's total
stock of heavy water from the condemned cell to a waiting vehicle.

The evacuation was to be through the port of Bordeaux. Here Halban found
the {\it Broompark}, a British coaler (fig.\ref{fig:broompark}). They 
were met by the Earl of
Suffolk,
liaison officer in France for the British Department of Scientific and
Industrial Research, who was charged with rescuing rare machine tools,
\$10M worth of industrial diamonds, fifty French scientists, and the heavy
water. The cargo was strapped to pallets on deck which would float free in
the event of the ship being sunk, and thus make rescue a possibility.
Joliot chose to remain in France and began a difficult period in charge of
the Coll\`ege de France cyclotron. This was the only one available to
German
Scientists in Occupied Europe, for although there was one in occupied
Copenhagen, Neils Bohr simply forbade its use by Axis personnel; an act
which only Bohr could conceivably have gotten away with\cite{26}. Later in
the
war Joliot went underground and became a leader of the Resistance.

Meanwhile in Bordeaux, the harbour was bombed, and the {\it Broompark}
sailed
down the Gironde estuary amid chaos. The ship next to her was sunk by a
mine; when questioned about the heavy water, Joliot said it was on this
ship. Eventually the heavy water reached London, where it was deposited
in Wormwood Scrubs prison. It was later moved to the Cavendish Laboratory
in Cambridge, where the Coll\`ege de France team were setting up to
continue
their experiments.

\section{World War II: The Western Allies}

A team coalesced in Cambridge around Hans von Halban and Lew Kowarski. By
1941 their experiments with uranium oxide and the 185kg of heavy water had
shown sufficient increase in neutron and fission activity to predict that
with 3-6 tons of heavy water, a self-sustaining chain reaction could be
achieved\cite{27}. A plan for industrial production of heavy water in
Britain by Imperial Chemical Industries (I.C.I.) was shelved in favour of an approach to the United States for
supplies. However, once the United States joined hostilities and the
centre of gravity of fission research moved inexorably westwards, it was
decided that the Cambridge team should move across the Atlantic. Chicago
was the first choice as that was the site of Compton's ``Met Lab", but this was
not possible due to American concerns about citizenship (some of the
Cambridge team were citizens of Axis nations) and ties with I.C.I. (a
rival chemical conglomerate to Du Pont, chief industrial partner to the
American effort)15.

Canada was a workable alternative and had the advantage that some fission
work had already been initiated there by Laurence and Sargent. Montreal
was chosen and in early 1943, the team, families, equipment, uranium and
heavy water made the journey while the Battle of the Atlantic was raging.
Once in Montreal, they were to interact freely, at least in theory, with
Met Lab team in Chicago. Detailed theoretical work on a prototype heavy
water/uranium reactor was to have begun immediately, following the work of
Fermi and Wigner on a carbon moderated reactor\cite{28}, but was delayed
by
worsening relations with the Americans. When it was finally started, after
a series of visits to Chicago by George Volkoff and George Placek in
January 1944, the American design effort was well underway.

The Americans had noted the British request for a heavy water supply in
1941, and Urey and Hugh Taylor of Princeton had suggested the Consolidated
Mining and Smelting Company of Trail as a likely source\cite{9}. The U.S.
National Defense Research Committee offered the company \$10 a pound for
high grade heavy water; at the time the American product cost \$1,130 a
pound\cite{27}.

In July 1942 it was clear the cost of Project No.9, as it was known, was
going to be enormous, but the U.S. government agreed to pay for it, with
the acquiescence of C. D.  Howe, the Canadian Minister for Munitions and
Supply. The Canadian Government only officially learnt of the project in
August 1942 \cite{16}. The final cost was \$2.8M, and the annual operating
costs
about \$700K. The company produced the heavy water for the American
nuclear
effort at cost, and did so from 1944 until 1956 when the plant became
uneconomical compared to the large U.S. plants which were by then in
production\cite{29}.

Cominco had produced electrolytic hydrogen since 1930 in a \$10M plant
consisting of 3,215 cells consuming 75MW of hydro-electric power. In
addition a tower was built for initial concentration (0.015\% to 2.3\%) by
D/H exchange on a catalyst; Hugh Taylor developed a platinum on carbon
catalyst for the first three stages and Urey developed a nickel-chromia
one for the fourth stage tower. The ``P-9 Tower" building housing this
plant still stands (figs.\ref{fig:trail_p1},\ref{fig:trail}). To the
electrolysis plant were added
secondary electrolysis cells used to increase the deuterium concentration
in the water from the exchange process from 2.3\% to 99.8\%
(fig.\ref{fig:trail_p2}, see 
appendix).
 
In November 1942, C. J. Mackenzie of the NRC requested that the Montreal
group receive the first year's output from Trail, about 6 tons. A rift then developed between the two groups, and in the late summer of 1943,
the Chicago group were given permission to build their own heavy water
pile at Argonne. Trail heavy water production started in January 1944 and
was sufficient for the Argonne pile - the first heavy water/natural
uranium reactor - to go critical on 15 May of that year. The project was
directed by Walter Zinn, a native of Kitchener, Ontario. In fact, the
Americans had decided on a full-scale program to produce plutonium with a
heavy water reactor, and DuPont eventually built three additional plants
on their own soil, in West Virginia, Indiana and Alabama.  These used the
fractional distillation technique; this method is very simple but very
heavy on power (see Appendix).

In April 1944 the rift between the Montreal group and the Americans was
patched up. American assistance, partly in the form of Trail heavy water,
was assured for the construction of a Canadian pile. The chosen site was
Chalk River. The first director of the new laboratory was to be John
Cockroft, although he was replaced before the first pile went critical by
W. B. Lewis, who went on to dominate the Canadian nuclear scene for
decades. Both men had a pre-war background of nuclear physics and wartime
experience of radar research and development in Britain. The pilot pile
was known as the NRX, and the 10 MW design was completed in July 1944.
However, it was not to be the first Canadian reactor. Cockroft proposed
building a smaller zero-energy pile making use of American experience at
Argonne\cite{17}. Thus, ZEEP (Zero-energy experimental pile)  was built, by
Section 6 of Chalk River's Nuclear Physics Division; Kowarski was Section
Head and B. W. Sargent was Acting Division Head\cite{16}. Criticality was
achieved in September 1945. NRX was completed and went critical in July
1947.

In France in September 1945 Allier obtained a promise from Norway that
France would receive the first five tons of heavy water produced once the
Norsk Hydro plant got started again. The original 185~kg, on loan since
1940 and still in use in Canada, was finally paid for. After ZEEP,
Kowarski returned to France and built that country's first pile, ZOE
({\it Zero
Oxide Eau Lourde}) in the old prison at Ch\^atillon. Criticality was
achieved
on 15th December 1948\cite{32}.

As a footnote to the history of the 185~kg of heavy water, the Norsk
Industriarbeidermuseum in Vemork would like to track down one of Joliot's
original containers for its exhibition.  However, after extensive
enquiries, it seems likely that these were disposed of when the Montreal
lab closed and was moved to Chalk River in 1947.

\section{World War II: Germany}

In contrast to the Allied effort, heavy water was central to Heisenberg's
rather ambiguous fission program in Germany\cite{33}. A large-scale
industrial
effort like that mounted in the United States for isotope separation was
impossible, and the entire program centred on a small amount of unenriched
uranium and Norwegian heavy water.

In February 1940 Werner Heisenberg, in Berlin, reviewed the published
literature, particularly the work of Bohr and Wheeler which pointed out
that it was the likely the rare isotope of uranium $^{235}$U which was
being
fissioned. He decided that to build a reactor from natural uranium
required good moderator - \hw\ or carbon. If \lw\ was to be used as a
moderator, then the uranium would have to be enriched in $^{235}$U
\cite{34}.
By
summer Walter Bothe, in Heidelberg, had tried the purest form of carbon he
could obtain - electrographite from Siemens - but found the absorption too
high for natural uranium. They recognized that this was probably due to
residual impurities, and worked out further refining steps, but Army
Ordnance, their paymasters, ruled out trials due to cost. Robert D\"opel
in
Leipzig had shown experimentally that \hw\ was a good moderator; indeed it
looked like the only possibility.

There was much experience with \hw\ in Germany; Paul Harteck, now at
Hamburg, had studied it with Rutherford, Munich physical chemist Klaus
Clusius had worked with it, and Bonh\"offer had worked with Norsk Hydro
from
the start. In a discussion as to the best method of producing the
material, Bonh\"offer favoured electrolysis, while Harteck thought
catalysis
more cost effective. They contacted the chemical conglomerate I. G. Farben
whom they hoped could handle production. External events took a hand in
April 1940, when German forces invaded Norway. After the situation was
stabilized - the Telemark region was the last part of Norway to be subdued
- I. G. Farben was given control of Norsk Hydro. Soon production with the
existing plant was increased from 20 litres per year to 1 ton per year.

As experimental work in Leipzig progressed ever more accurate estimates
were made of the amount of heavy water required by an uranbrenner (uranium
burner, i.e. reactor). This was 4-5 tons, and Harteck and Suess developed
a catalytic exchange method (dual temperature H/\lw , essentially the
``Trail Method") for raising production to the level of one reactor
charge
per year. In 1941 the Norwegians were forced to install a catalytic plant at
their own cost, as their contribution to the Axis war effort. At the end
of 1941, possibly in anticipation of the trouble to come, Harteck and
Clusius proposed a new plant in Germany. He surveyed the possibilities:
the two used in Rjukan, and repeated distillation of hydrogen. However,
Army Ordnance preferred Norway as the cheapest option.

Meanwhile, experimental and theoretical work on reactor design was
proceeding apace. An alternating layer design was produced by Heisenberg
and Weizs\"acker's graduate students Karl-Heinz H\"ocker and Paul
M\"uller. In
addition to D\"opel working with uranium metal and heavy water in Leipzig,
Karl Wirtz at Berlin-Dahlem did experiments with uranium oxide and
paraffin. It looked increasingly like \hw\ would work. In addition,
Weizs\"acker started exploring the explosive possibilities of element 94,
which would inevitably be produced in the uranbrenner. Despite this, in
early 1942, Army Ordnance decided fission was irrelevant for the war
effort and proposed suspending research.

By the end of 1942, the picture was mixed. The first neutron increase
(13\%) in a heavy water/uranium pile, the L-IV, had been observed by D\"opel 
and Heisenberg in Leipzig. However, the catalytic plant in Rjukan was
still under construction so heavy water was still in short supply. In
addition there had been an attempted commando raid on Rjukan; it had been
a disastrous failure and the fate of the commandos makes unpleasant
reading\cite{35}, but it was a sign of things to come.
  
In a second commando operation, on 28th February 1943, Norwegian saboteurs
guided by Tronstad (in London) and Brun (locally) destroyed the Rjukan
electrolysis plant, with the loss of 500kg of heavy water. This highly
economical and clinical operation has long been celebrated in Norway,
particularly in the Rjukan area, where residents retrace each year on July
1 (note not in February!) the steps of the commandos down from the
Hardanger Plateau where they gathered prior to the attack. In a war that
killed maybe 50 million people, this courageous action, which resulted in
no injuries and, mercifully, no reprisals, is well worth remembering.

The plant was, however, quickly repaired and General Groves, fulminating
against British control of Norwegian special operations, ordered the job
to be done ``properly": by massive air bombardment. A major air raid by the USAAF followed on 16th
November 1943 and resulted in considerable loss of civilian life and
minimal damage to the electrolysis building. Clearly the Allies thought
nuclear fission worth pursuing, even if the German Army didn't. It was
decided to ship all the heavy water and its production to Germany, mostly
because of the fear that further raids might endanger explosives
production at Vemork. On 20th February 1944, the last shipment of heavy
water from Rjukan to Germany was lost when the ferry carrying it across
Lake Tinnsj\"o was sabotaged. This was the last time Rjukan people died
in what Abraham Esau described as the ``Norwegian National Sport" of
destroying heavy water plant. Norsk Hydro received no payment for the
heavy water produced, and had suffered 16 million kroner of damage to the
plant.

The effort to build a plant in Germany dragged on for the next year, mired
in bitter price and patent disputes with I. G. Farben. The failure to produce more heavy water caused severe difficulties for the experimental groups,
and air dilution was downgrading existing stocks. Attempts were made to
build a polishing plant to upgrade exisiting stocks using equipment looted from Norway. Air attacks on
Hamburg and Berlin forced the evacuation of research labs to the
countryside.

Heisenberg and his team moved to Haigerloch, near the Swiss Border, where
they hoped they could work in relative peace. The laboratory was
constructed in the side of a cliff beneath an old church. They were still
trying to achieve criticality with the B-VIII reactor when first contact
was made with American forces: the fact-finding ``Alsos" group\cite{36}.
With
lingering worries about the German nuclear threat, and always mindful of
post-war commercial competition and the fact that Haigerloch was to be in
the French zone, an American army team dynamited the laboratory before
handing the region over to their allies, who were just then entering in
the town.

Why did they fail? There are many reasons as the reader will have
gathered, but perhaps the most pivotal is that there were never more than
70 scientists involved, and only 40 worked more than half time on the
uranbrenner. (This is in stark contrast with the \$2 billion Manhattan
Project). Only a handful of these scientists were of the first rank;
Germany had lost its best to emigration, and of those that remained the
most brilliant worked in aerodynamics and rocketry.

Although never used in Germany, Karl-Hermann Geib in Leuna in 1943
developed what we now regard as the most cost-effective process for
producing heavy water: the dual temperature exchange sulphide process (see 
appendix and fig. \ref{fig:sulphide}). Contemporaneously, the process was
also developed by J. S.
Spevack at Columbia University\cite{37}, and his process became the basis
of the
post-war North American plants under the name of the Girdler Process,
named after the company which first exploited it. North American
scientists were not aware of Geib's work for many years after the war;
Maloney et al. in their book ``The Production of Heavy Water"
(1955)\cite{38}
complain that relevant German wartime work was still classified.

Unfortunately Geib was not able to benefit from his work; in 1945 he was
taken to the USSR, along with many others, who were given a 10-year
contract to work on fission and aerodynamics. Many German scientists found
this very congenial and some even went as far as to describe these 10
years as the time of their lives. However, Geib was not so happy and he
made the mistake of applying for asylum in Canada, giving the name of
Professor E. W. R. Steacie as a reference. Officials at the Canadian
Embassy in Moscow did not know what to do with him and told him to come
back the next day. That was the last time he was seen. His wife in
Germany received his effects in the mail\cite{33}.

\section{World War II: The Soviet Union}

In October 1939 Zel'dovich and Khariton reviewed Joliot's and Fermi's
results, and concluded, correctly, that heavy water and carbon were the
only feasible moderators for a natural uranium reactor\cite{39}.

In August of 1940 Kurchatov, Khariton, Flerov and Rusinov submitted a plan
to the academy ``On the utilization of the energy from U fission in a
chain
reaction". Khariton calculated 2.5t of uranium oxide and 15t of heavy
water were needed for a reactor. The shortage of uranium was the main
problem for experimenters; ironically the Soviet Union, in spite of its
huge size, had no uranium mines. Kurchatov sent young colleagues to
Leningrad photographic shops to buy uranium nitrate.

In 1941, German forces invaded the Soviet Union, and Flerov found himself
in the Red Air Force near Voronezh. The previous year he and Petzhak had
discovered the spontaneous fission of uranium and he was interested to see
if there had been any reaction to this event from his colleagues abroad.
Looking in the local university library he found that all the scientists
he knew to be working on fission had disappeared from the literature. This
was the first clue Soviet scientists had of what was going on in the West.
Flerov sent a letter to Stalin, warning him. Soon a concerted Soviet
fission effort was initiated, with Kurchatov in charge.
 
By March 1943 Kurchatov was receiving Fuchs' reports and was abreast of
work in the USA. At this time there was only 2-3kg of heavy water in
country, and partly because of this, and because General Groves had
restricted heavy water exports from the United States, they chose graphite
for their first reactor. More uranium was needed than with heavy water but
electrolytic graphite was available in the country. A heavy water plant
had been planned before the war at a nitrogen plant in Chirchik,
Tajikistan, but it had not been completed. Alikhanov was given the more
distant task of building a heavy water pile, as he would not work under
Kurchatov and this was a convenient way of removing him from the main
group. Construction began after the war.

In late 1943 the Soviet purchasing commission in the U.S. obtained 1~kg of
heavy water and a further 100~g in February 1945. Given the munificence of
these gifts, it seems as if the early decision to concentrate on graphite
was a good one! On December 25th, 1946 the graphite reactor F-1 went
critical.

Work on the Chirchik Nitrogen Combine plant started in 1944; the Institute
of Physical Chemistry in Moscow started research on the physical chemistry
of electrolytic cells at the same time. After hostilities ceased in
Europe, the NKVD assembled German heavy water experts at the Leuna works
in Merseberg, where they designed a new heavy water plant. In October 1946
they were moved to the Institute of Physical Chemistry in Moscow. The plant
producing large quantities by 1948. Alikhanov's reactor at the
Thermotechnical Institute in Moscow went critical in April 1949.

\section{Postscript}

Since World War II only Canada amongst the industrialized nations has
pursued the commercial possibilities of heavy water/natural uranium
reactors. A large number of CANDU (Canadian-Deuterium-Uranium) reactors
are now operating in Eastern Canada, and the reactor and heavy water
technology has been exported to several other countries.  A downturn in
the nuclear industry in the 1980s led to one thousand tonnes of heavy
water being made available for use as a neutrino detector in the Sudbury
Neutrino Observatory, which is now (2011) operational in Northern
Ontario. I was a member of this effort from 1988-2009, which is how I became interested in
the subject of this paper.

\section*{Acknowledgments}

The author would like to thank Cominco personnel Richard Fish and
Brent Cross for their assistance regarding the history of Project 9 and a
fascinating tour of the derelict Trail heavy water tower in 1995. One of
the photographs is reproduced with permission from the Cominco Magazine. I
thank Norsk Hydro chemist and historian Per Pynten for a marvellous day at
Vemork and Rjukan in 1996. I am indebted to Professor Maurice Pryce for
pointing me toward several valuable references, and to the late George
Volkoff for reading an early version of this manuscript.

Many thanks to Don Cody of California, grandson of the Broompark's skipper Captain Paulsen, for 
information about, and the photograph of, the Broompark.

\section*{Afterword}

This article was written between 1992 and 1998. In 1999 Per Dahl's book
``Heavy Water and the Wartime Race for Nuclear Energy" was published. Dahl 
covers many aspects of this story in great detail and the book is highly
recommended to interested readers.

\section{Appendix}

\subsection{Separation of Deuterium and Heavy Water}

There are four basic methods of producing heavy water, they have different
economics depending on the initial and final concentrations, and so they
tend to used in combination.

1.  Distillation of water or hydrogen. Heavy water boils at 101.4C, only
slightly higher than light water, so the process requires many stages and
a large amount of thermal energy. The separation effect is bigger for
liquid hydrogen, but this is less convenient to work with than water.

2.  Electrolysis of water. H$^+$ ions are evolved preferentially over
D$^+$. As
above, multi-stage and heavy on electrical energy, now used as final
``polish".

3.  Hydrogen/water catalytic exchange, the ``Trail Method", depends on the
fact that, when isotopic equilibrium is established between hydrogen gas
and water, the water contains 3 or 4 times as much deuterium as does the
gas. See figure \ref{fig:trail_method}.

\beq
\mbox{H}_2\mbox{O} + \mbox{HD} \rightarrow \mbox{HDO} + \mbox{H}_2
\eeq

This method can also be used in dual-temperature mode (see point 4 below)

4. The Dual-Temperature Sulphide Process relies on the following reaction:

\beq
\mbox{H}_2\mbox{O}(liquid) +  \mbox{HDS}(gas)  \rightarrow
\mbox{HDO}(liquid) + \mbox{H}_2\mbox{S}(gas)
\eeq

Deuterium enrichment in the water increases as the temperature decreases.
This method has two big advantages over the Trail method. The use of
hydrogen sulphide eliminates the need for a catalyst, and the dual
temperature aspect means that the hydrogen sulphide is maintained in a
closed loop (see figure \ref{fig:sulphide}) and no electrolytic
regeneration of hydrogen is
required.

Modern Practice (Bruce Plant, Ontario)

* Three-stage dual temperature sulphide process (0.015\% - 25\%)

* Distillation with waste reactor heat (25\% - 99.85\% reactor grade)

\subsection{Neutron-Induced Fission}

A slow neutron is absorbed by a  $^{235}$U nucleus, which splits into two
medium-mass nuclei, emitting 200 MeV of energy and, on average, 2.5 fast
neutrons.
For example:

\beq
\mbox{n} + ^{235}_{92}\mbox{U} \rightarrow  ^{93}_{37}\mbox{Rb} +
^{141}_{55}\mbox{Cs} + 2\mbox{n}
\eeq

To promote a chain reaction, the outgoing fast neutrons have to be slowed
by collision with nuclei in a moderator, without absorption.

\subsection{Moderators}

The best moderator would be hydrogen, whose nucleus has a mass equal that
of the neutron; all the neutron's energy can be lost in one collision.
However, hydrogen absorbs neutrons, producing a deuterium nucleus (D; or
$^2$H ) and a gamma-ray:

\beq
\mbox{n + p} \rightarrow \mbox{D} + \gamma
\eeq

In reality the best moderator is
deuterium, which cannot
easily absorb a second neutron. The most convenient form of deuterium is
Heavy Water (\hw ).

Carbon also makes a good moderator, but tends to have neutron-absorbing
impurities; it therefore has to be very pure (Electrolytic Graphite).

\newpage

\section{Figures}

Fig 1: Author at hydro-electric generator building at Vemork, just outside
Rjukan,
now the Norsk Industriarbeidermuseum (Industrial Workers Museum). The
heavy water plant was in a (rather unattractive) building, now demolished,
in front of this one. (Author photo).

Fig 2: Part of the electrolysis plant from Rjukan, now reassembled in
the Resistance
Museum, Oslo. The longer steel cylinders are for electrolysis and the
shorter ones are the combustion chambers. The brown tubes above are
condensers.  (Author photo).

Fig 3: Norsk Hydro heavy water order book, pages 1 and 3. (Per Pynten).

Fig 4: The Broompark sailing out of Le Verdon Roads in 1940
with the supply of heavy water, under the command of Captain Olaf Paulsen.
The Broompark was built in 1939 and
torpedoed and sunk in 1942. The Denholm Shipping Lines office in
Glasgow have a glass-encased six-foot model of her. (Don Cody)

Fig 5: The Trail Heavy Water tower in the 1940s. (B.C. Archives and
Records Service Cat. No. NW669 C733 V.30-32/1). 

Fig 6: The Trail Heavy Water tower in 1995. (Author photo).

Fig 7: A few of the secondary electrolytic cells at the Trail Plant
(B.C. Archives and Records Service Cat. No. NW669 C733 V.30-32/1).

Fig 8: One of the containers used for shipping Trail heavy water, now in
the museum at Rossland, B.C. (Author photo).

Fig 9: Trail method for \hw\ enrichment.

Fig 10: Sulphide method for \hw\ enrichment.

\newpage

\newpage
\begin{figure}
\center{\epsfig{file=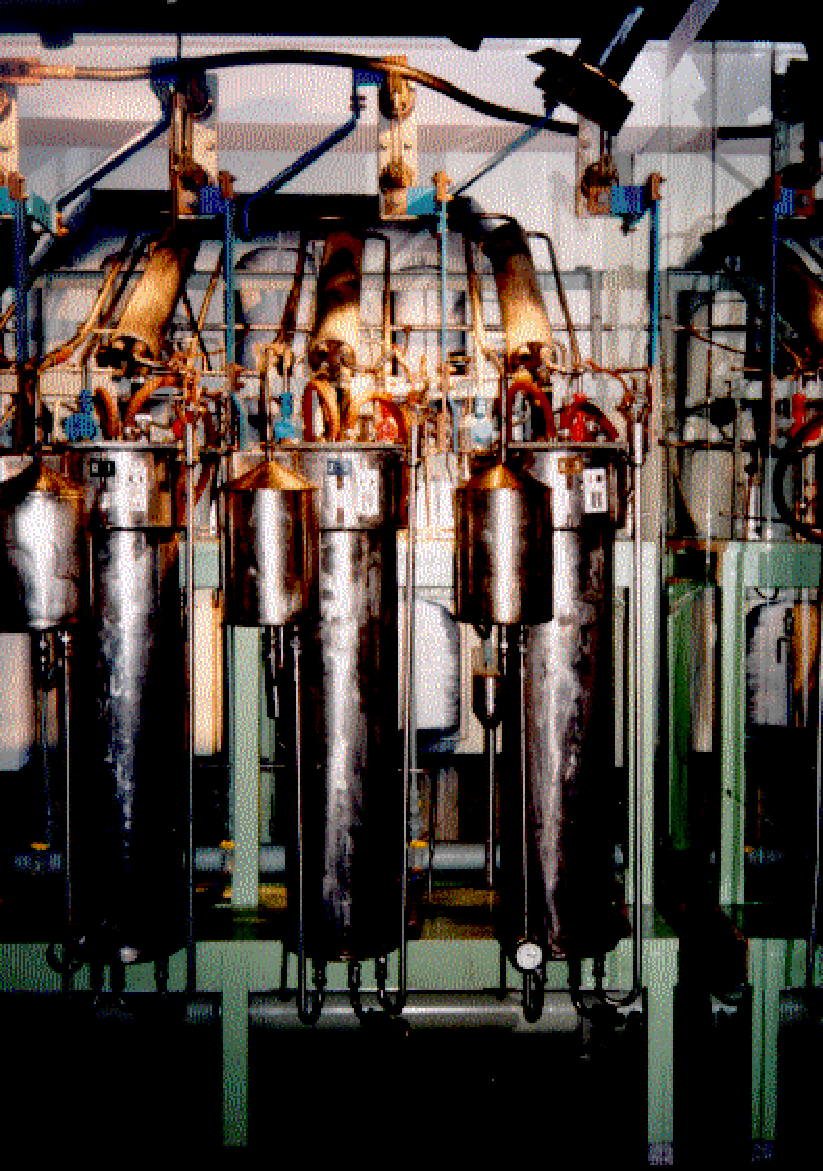,width=10cm}}
\caption{Part of the electrolysis plant from Rjukan, now reassembled in
the Resistance
Museum, Oslo. The longer steel cylinders are for electrolysis and the
shorter ones are the combustion chambers. The brown tubes above are
condensers.  (Author photo).}
\label{fig:electrolysis}
\end{figure}

\newpage
\begin{figure}
\center{\epsfig{file=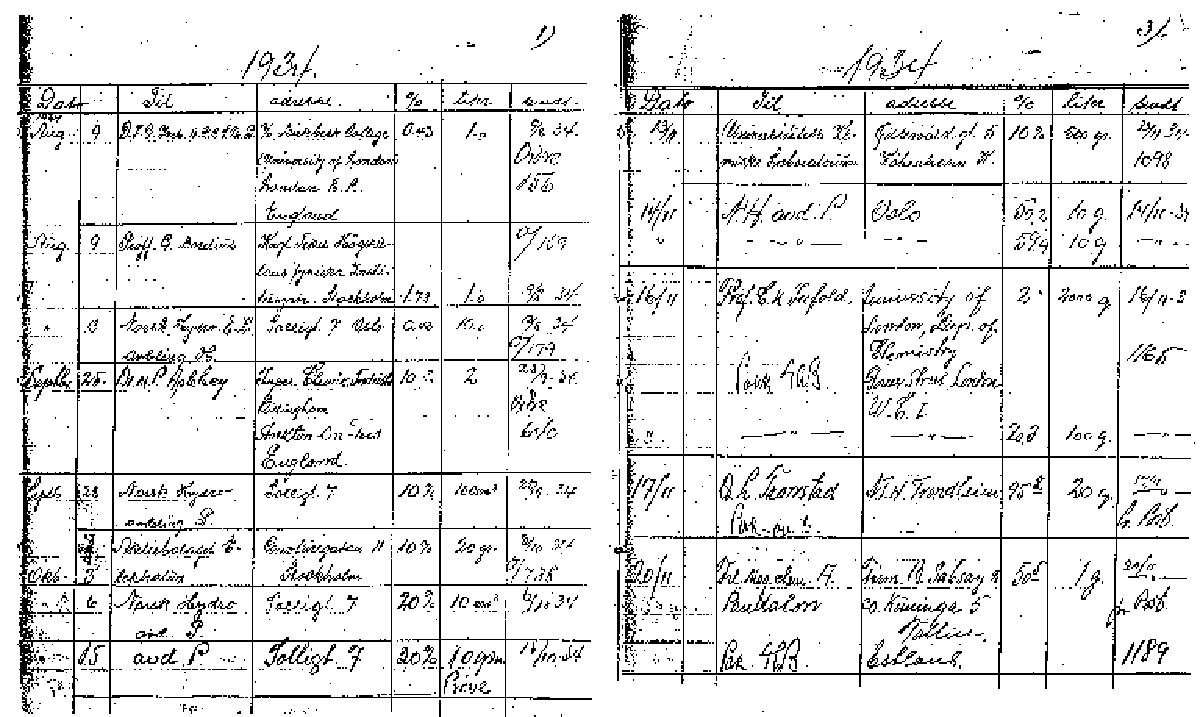,width=15cm}}
\caption{Norsk Hydro heavy water order book, pages 1 and 3. (Per Pynten)}
\label{fig:order}
\end{figure}

\newpage
\begin{figure}
\center{\epsfig{file=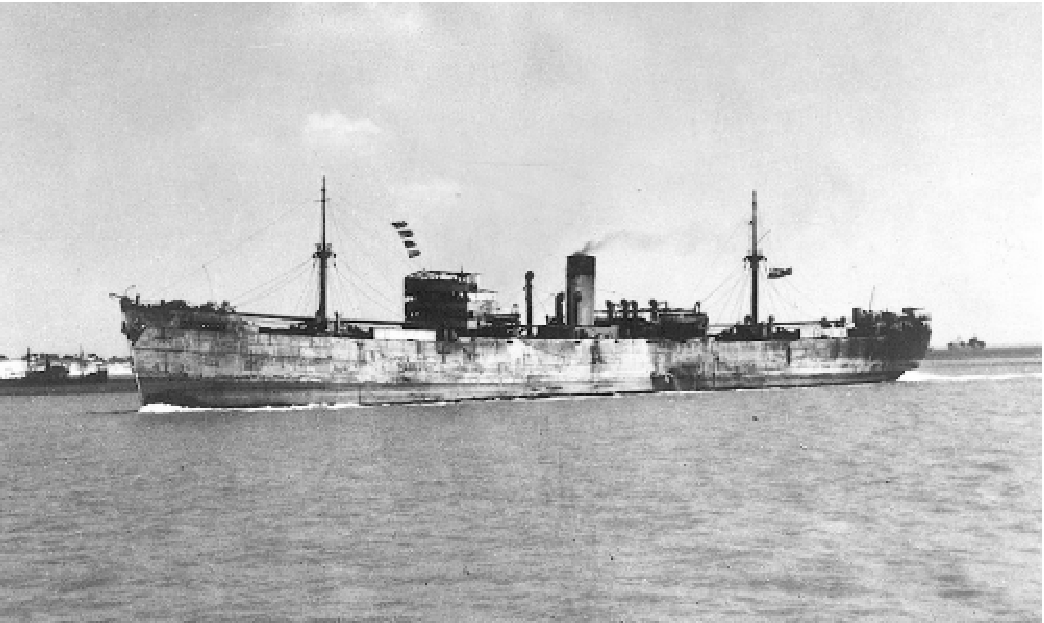,width=15cm}}
\caption{The Broompark sailing out of Le Verdon Roads in 1940 
with the supply of heavy water, under the command of Captain Olaf Paulsen. 
The Broompark was built in 1939 and 
torpedoed and sunk in 1942. The Denholm Shipping Lines office in 
Glasgow have a glass-encased six-foot model of her. (Don Cody)}
\label{fig:broompark}
\end{figure}

\newpage
\begin{figure}
\center{\epsfig{file=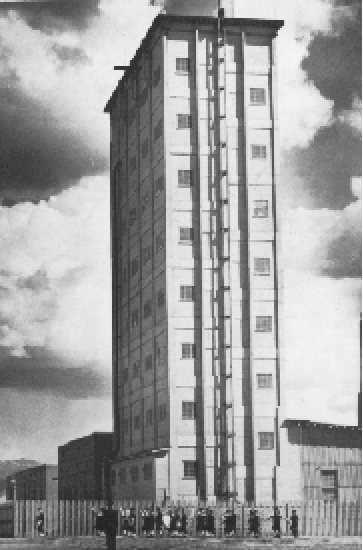,width=10cm}}
\caption{The Trail Heavy Water tower in the late 1940s
(B.C. Archives and Records Service Cat. No. NW669 C733 V.30-32/1)}
\label{fig:trail_p1}
\end{figure}

\newpage
\begin{figure}
\center{\epsfig{file=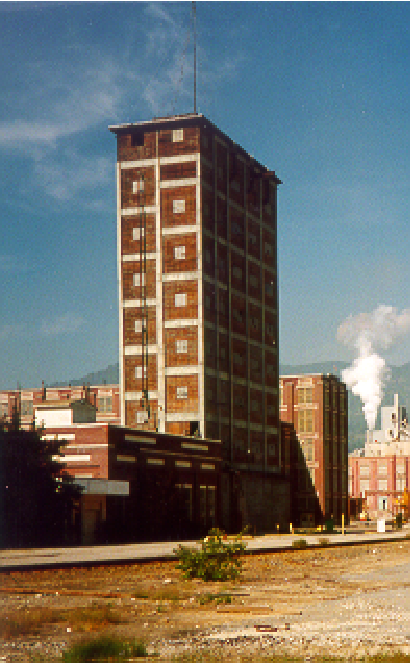,width=10cm}}
\caption{The Trail Heavy Water tower in 1995
(author photo)}
\label{fig:trail}
\end{figure}

\newpage
\begin{figure}
\center{\epsfig{file=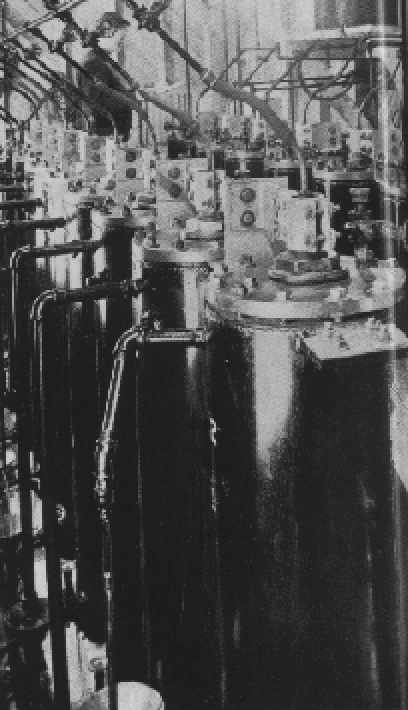,width=10cm}}
\caption{A few of the secondary electrolytic cells at the Trail Plant
(B.C. Archives and Records Service Cat. No. NW669 C733 V.30-32/1)}
\label{fig:trail_p2}
\end{figure}

\newpage
\begin{figure}
\center{\epsfig{file=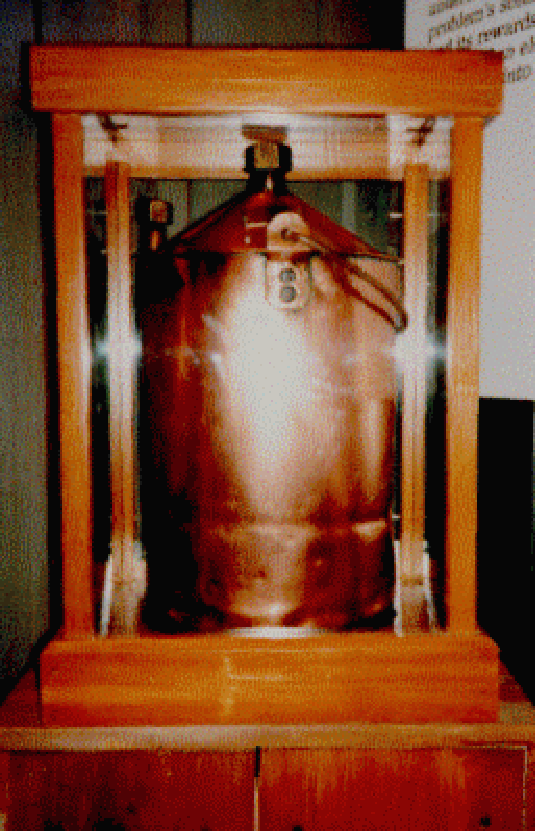,width=10cm}}
\caption{One of the containers used for shipping Trail heavy water, now in
the museum at Rossland, B.C. (Author photo)}
\label{fig:can}
\end{figure}

\newpage
\begin{figure}
\center{\epsfig{file=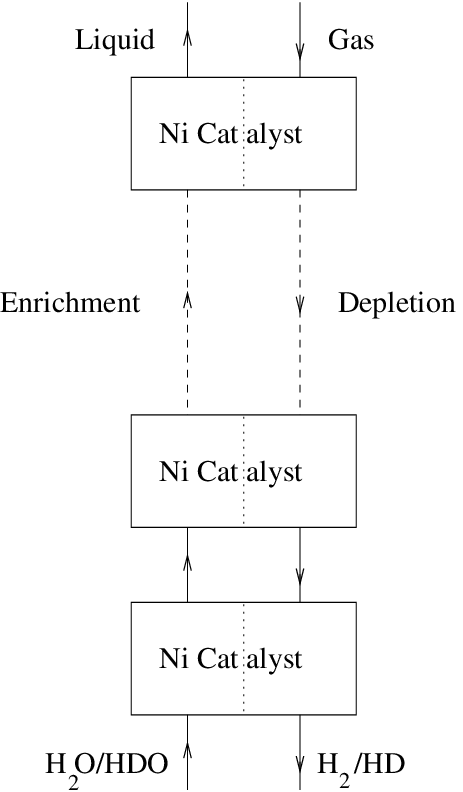,width=10cm}}
\label{fig:trail_method}
\caption{Trail method for \hw\ enrichment}
\end{figure}

\newpage
\begin{figure}
\center{\epsfig{file=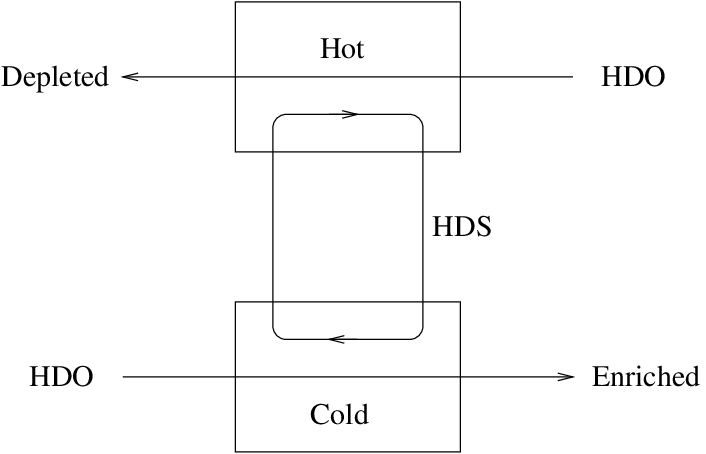,width=10cm}}
\caption{Sulphide method for \hw\ enrichment}
\label{fig:sulphide}
\end{figure}

\end{document}